\documentclass[usenatbib,usedcolum,usegraphicx]{mn2e}
\title[
OGLE small amplitude red giant variables in the Galactic Bar
      ]{
OGLE small amplitude red giant variables in the Galactic Bar
 }

\author[             J. J. Wray, L. Eyer \& B. Paczynski
       ]
       {             J. J. Wray,$^{1}$ L. Eyer$^{1,2}$\thanks{E-mail: laurent.eyer@obs.unige.ch}
                  \& B. Paczynski$^1$\\
  $^1$ Princeton University Observatory, Princeton, NJ 08544-1001, USA\\
  $^2$ Observatoire de Gen\`eve, CH-1290 Sauverny, Switzerland
}
\begin{document}
\maketitle
\label{firstpage}

\begin{abstract}
  Among over 200,000 Galactic Bulge variable stars in the public
  domain OGLE catalogue, we found over 15,000 red giant variables
  following two well defined period -- amplitude relations.  The
  periods are in the range $10 < P < 100$ days, and  amplitudes
  in the range $0.005 < A < 0.13$ mag in I-band.  The
  variables cover a broad range of reddening corrected colours,
  $1 < (V-I)_0 < 5$, and a fairly narrow range of extinction corrected
  apparent magnitudes, $ 10.5 < I_0 < 13 $.  A subset of variables
  (type A) has a rms scatter of only 0.44 mag.  The average magnitudes
  for these stars are well correlated with the Galactic longitude, and
  vary from
  $ I_{k,0} = 11.82 $ for $ l = +8^{\circ} $ to
  $ I_{k,0} = 12.07 $ for $ l = -5^{\circ} $,
  clearly indicating that they are located in the Galactic Bar.
  Most variables have several oscillation periods.

\end{abstract}


\section{Introduction}
While the existence of the Galactic Bar was pointed out decades ago
\citep{devaucouleurs64}, the Bar existence received a broader support
only following the work of \citet{blitz91}, based on 2.4 micron light
distribution in the inner Galaxy as measured by \citet{matsumoto82},
and the work of \citet{nakada91}, based on the asymmetry in the
distribution of Galactic Bulge stars as measured by IRAS.  The
presence of the Galactic Bar has been observed in many different ways;
recent reviews are provided by \citet{garzon99}, \citet{gerhard02},
\citet{dehnen02}, and \citet{merrifield03}.

The Optical Gravitational Lensing Experiment (OGLE) is a long term
observing project carried out at the Las Campanas Observatory in
Chile, operated by the Carnegie Institution of Washington.  The
instrumentation of the three stages of the project was described by
\citet{udalski92}, \citet{udalski97}, and \citet{udalski02a}.  A
catalogue of over 200,000 variable stars discovered within
approximately 11 square degrees in the direction of the Galactic Bulge
was published by \citet{wozniak01}.  The public domain data set
contains $ 200 - 400 $ I-band photometric data points obtained in the
observing seasons 1997, 1998, 1999 with the OGLE-II instrumentation: a
dedicated 1.3-meter telescope with a $ 2K \times 2K $ pixel CCD
camera, built and operated by the Warsaw University Observatory.

\begin{figure*}
\resizebox{\hsize}{!}{\includegraphics[angle=0]{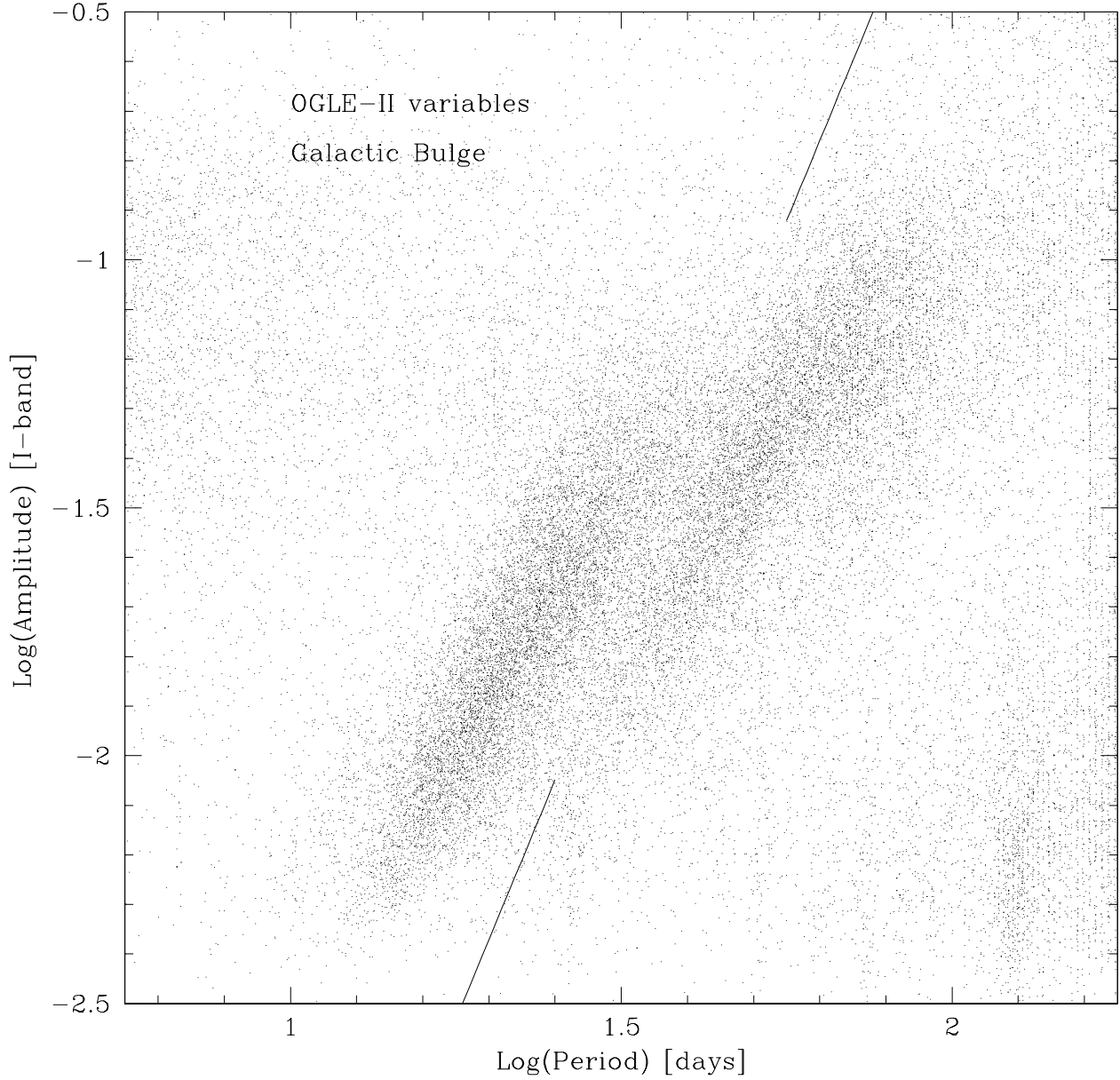}}
\caption{\label{fig1}
  The period -- amplitude diagram is shown for $ \sim 200,000 $
  OGLE-II variables in the Galactic Bulge.  Two sequences, separated
  by two segments of diagonal solid line, are clearly visible.  We call
  the sequence above and to the left of the line type A, and the sequence
  below and to the right of the line we call type B.  A cluster of points
  at $ \log P \approx 2.1 $ and $ \log A \approx -2.3 $ is an artefact to
  be ignored. Amplitudes are defined as half of the peak-to-peak
  amplitude for a given mode.
}
\end{figure*}

The presence of the Galactic Bar was already detected in the apparent
distribution of red clump giants in the OGLE-I data
(\citealt{stanek94}, \citealt{stanek97}).  In this paper, we set out to
determine whether the evidence for the Galactic Bar may also be found
among OGLE variables.  We calculated periodograms (\citealt{lomb76},
\citealt{press92}) for each of the 200,000 objects in the OGLE-II
variable star catalogue \citep{wozniak01} that had at least 100 I-band
measurements, and accepted as real those for which the probability of
a false signal \citep{press92} was smaller than 0.001.  Because fields
45 and 46 have only two seasons of observations (fewer than 100
measurements), periodograms were not calculated for any stars in these
fields.  We were struck by the two clear sequences apparent in the
period -- amplitude diagram, presented in Fig.~\ref{fig1}.  For many
stars, several significant periods were found.  The plotted period has
the largest amplitude.  Our convention is that amplitudes are defined
as half of the peak-to-peak amplitude for a given mode.

Originally, we had intended to use Mira-type long period variables as
possible tracers of the Galactic Bar.  However, it is clear that the
variables defining the two sequences in Fig.~\ref{fig1} are more
numerous and, having small amplitudes, they appear to be more useful
than Miras as tracers of the Galactic structure.  Therefore, this paper
is devoted to these low amplitude variables only.

In addition to the two sequences present in Fig.~\ref{fig1} there is a
clump of variables around $\log P \sim 2.1$ and $\log A < -2$; this is
mostly caused by an artefact, and should be ignored.

\begin{figure*}
\resizebox{\hsize}{!}{\includegraphics[angle=0]{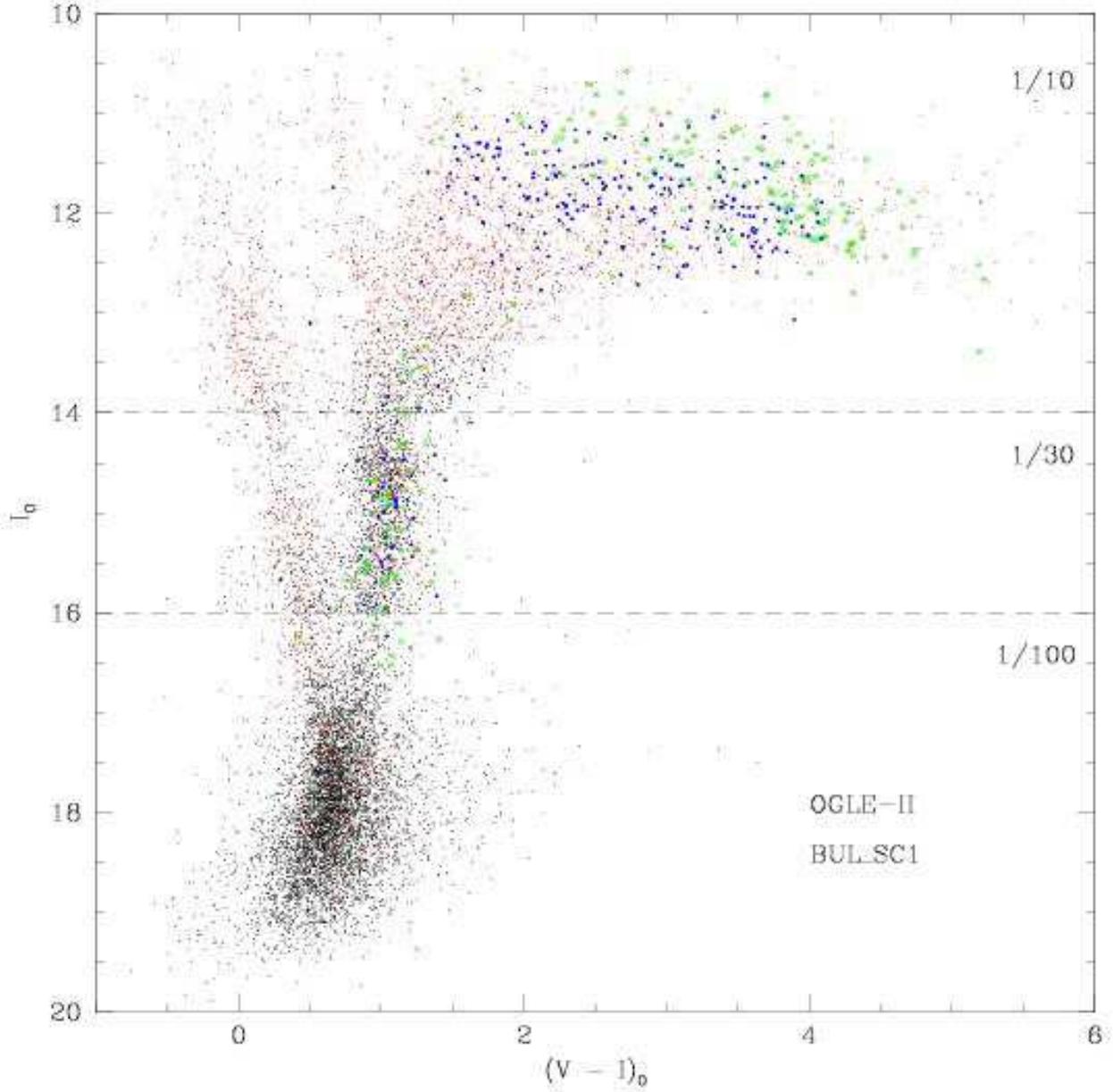}}
\caption{\label{fig2}
  The extinction and reddening corrected colour -- magnitude diagram
  $ I_0 - (V-I)_0 $ is shown for type A (large blue dots) and type B
  (green open circles) variables in the OGLE-II field $ {\rm
    BUL\_SC1} $.  Small red points represent other variables.  Small
  black points represent non-variable stars.  In the three sections of
  the diagram, separated by horizontal dashed lines, the fraction of
  non-variable stars plotted is as indicated: 1/10, 1/30, and 1/100, from
  top to bottom.
}
\end{figure*}

We divided the variables into types A and B, with the division line
shown in Fig.~\ref{fig1}.  Type A is above and to the left, and type
B below and to the right, of the division line.  Variables within the
parallelograms:
$$
\begin{array}{ccccl}
-2.31 & < & \log (A)                 & < & -1.31 ,     \\
-6.56 & < & \log (A) - 3.223\log (P) & < & -5.66 , 
\end{array}
\eqno(1a)
$$
$$
\begin{array}{ccccl}
-1.82 & < & \log (A)                 & < & -0.90 ,     \\
-7.62 & < & \log (A) - 3.223\log (P) & < & -6.56 , 
\end{array}
\eqno(1b)
$$
\noindent
were selected in 33 OGLE-II fields spanning a large range of Galactic
longitudes, but only a small range of Galactic latitudes.  The colour
-- magnitude diagram, corrected for interstellar reddening and
extinction with the maps provided by \citet{sumi03}, is shown in
Fig.~\ref{fig2} for the field BUL\_SC1.  Type A variables are marked
with large blue dots; B variables are marked with green open circles.
A relatively small range of apparent I-band magnitudes and a large
number of variables indicates that these stars might be useful
`standard candles,' and that it might be possible to use them as
tracers of the Galactic Bar.

Most variables shown in Fig.~\ref{fig2} are bright red giants.  It is
very likely that these stars are similar to the variables found by
\citet{kiss03} and \citet{ita03} in the OGLE catalogue for the LMC and
SMC, to those found by \citet{cook97}, \citet{wood99}, and
\citet{wood00} in the MACHO database for the LMC, to the M giant
variables in NGC 6522 \citep{glass03}, and to those found by
\citet{cook97} in the MACHO database for the Galactic Bulge.  Minimum
variability levels have been detected for red giants from the ESO
project ``long term photometry on variables" (\citealt{jorissen97}),
and also from Hipparcos photometry (\citealt{eyer97}), the cooler
stars having larger minimum variability thresholds and larger mean
amplitudes.  Furthermore, a period-amplitude relation has been noticed
for a few red giants stars by \citet{jorissen97}, and also in a large
sample of stars from ASAS data (\citealt{eyer02}).  A possible
theoretical mechanism for the oscillations was studied by
\citet{dziembowski01} and \citet{christensen01}.

The aim of our paper is to select from the OGLE catalogue red giant
variables populating the two period -- amplitude sequences in
Fig.~\ref{fig1}, to provide many parameters for these variables in
electronically accessible tables, and to check if they trace the
Galactic Bar.

\section{Selection of variables}
\begin{figure*}
\resizebox{\hsize}{!}{\includegraphics[angle=0]{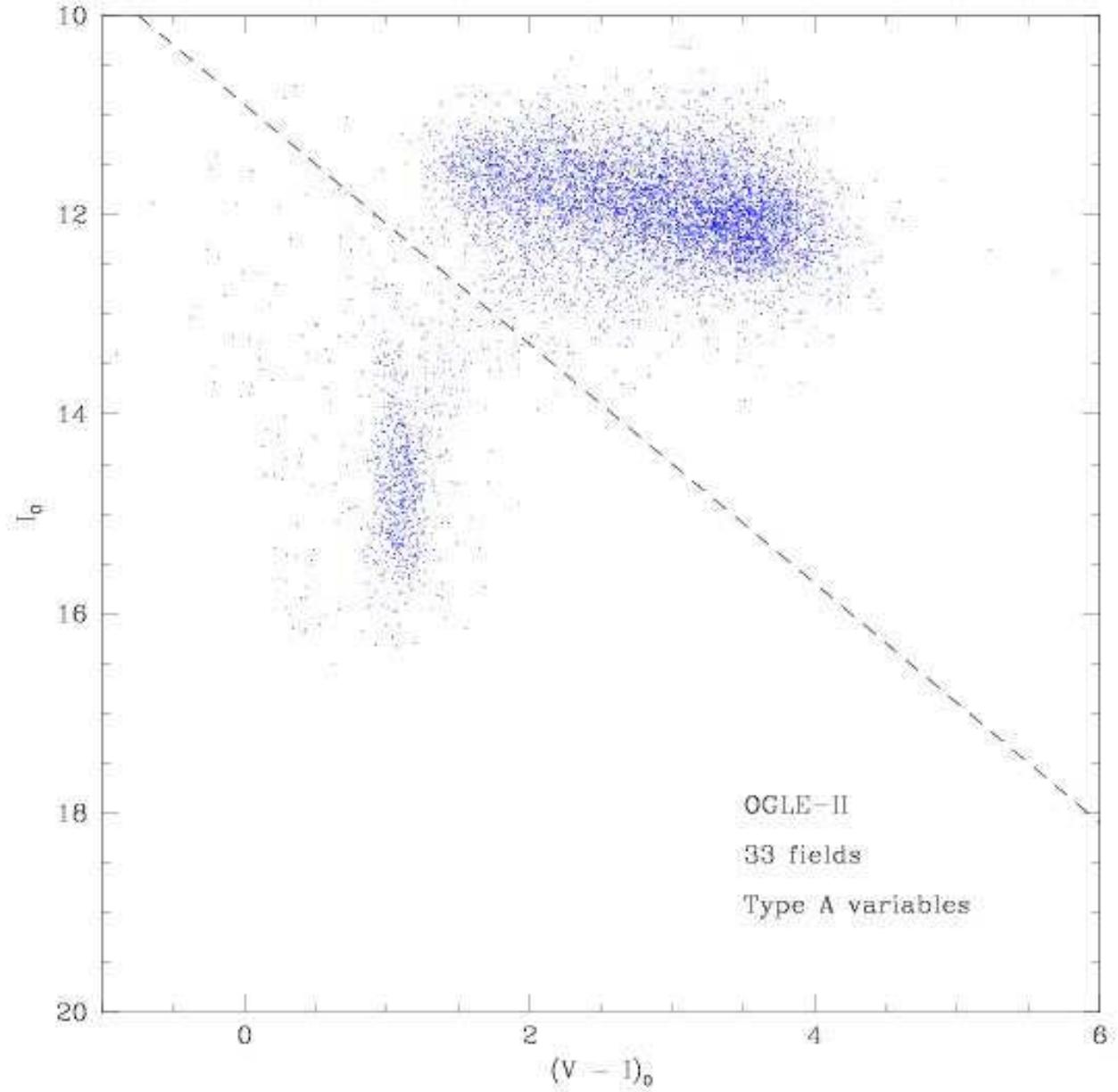}}
\caption{\label{fig3}
  The extinction and reddening corrected colour -- magnitude diagram
  $ I_0 - (V-I)_0 $ is shown for type A variables in 33 OGLE-II fields.
  The variables to be discussed in this paper are located above and to the
  right of the diagonal dashed line.
}
\end{figure*}

We used reddening maps published by \citet{sumi03} for 48 OGLE-II
Galactic Bulge fields.  All variables of type A in a subset of these
fields located at the Galactic latitude $ b \approx -4^{\circ} $ are
shown in Fig.~\ref{fig3}.  It is clear that they form three distinct
groups: luminous red giants, less luminous red giants located near the
Bulge red clump region, and relatively blue Galactic disc stars.
These three groups are well separated.  We selected stars located
above and to the right of the line in Fig.~\ref{fig3} for our study,
and we refer to them as OGLE Small Amplitude Red Giant variables
(OSARGs).

Some type B variables are close to the saturation limit of OGLE-II, and
therefore we restrict our study of the Galactic structure
to OSARG type A variables only.  The OGLE saturation limit causes a
problem for the study of the brightest OSARGs.  This technical difficulty
will be overturned when the ASAS I-band data \citep{pojmanski02} become
available, which have a limiting I-mag of about 14, and a considerable
overlap with OGLE-II photometry. Three years of Galactic Bulge data have
been acquired by ASAS already.  For our analysis we omit fields located
far from $ b \approx -4^{\circ} $, and those fields for which a
reliable extinction map is not available.  This leaves us with a total
of 33 OGLE-II fields.

\section{Distribution of variables}
The OSARG type A variables form a nearly horizontal sequence in the
colour -- magnitude diagram shown in Fig.~\ref{fig3}.  We fitted
their distribution with a straight line using the least square method,
and looking for the best global solution of the form:
$$
I_{k,i,0} = I_{k,0} + a \left[ (V-I)_{k,i,0} - 2.881 \right] ,
\eqno(2)
$$
where $ I_{k,i,0} $ and $ (V-I)_{k,i,0} $ are the reddening
corrected average magnitude and colour for variable number $i$ in the
field $k$, $I_{k,0}$ is a constant, different for every field,
representing the `average' magnitude of OSARG type A variables in
field $k$, $a$ is the same constant for all fields, and $2.881$ is the
mean $ (V-I)_0 $ colour for all OSARG type A variables in all 33
fields represented in Fig.~\ref{fig3}.  We computed the value of the
parameter $a$ as 0.169.  The values of the parameter $ I_{k,0} $ and
the standard deviations from the relation (2) are given in
Table~\ref{tab1} for all 33 fields.  This table also lists the
total number of type A and type B variables in each field and the
Galactic coordinates of the field centers.

\begin{figure*}
\resizebox{\hsize}{!}{\includegraphics[angle=0]{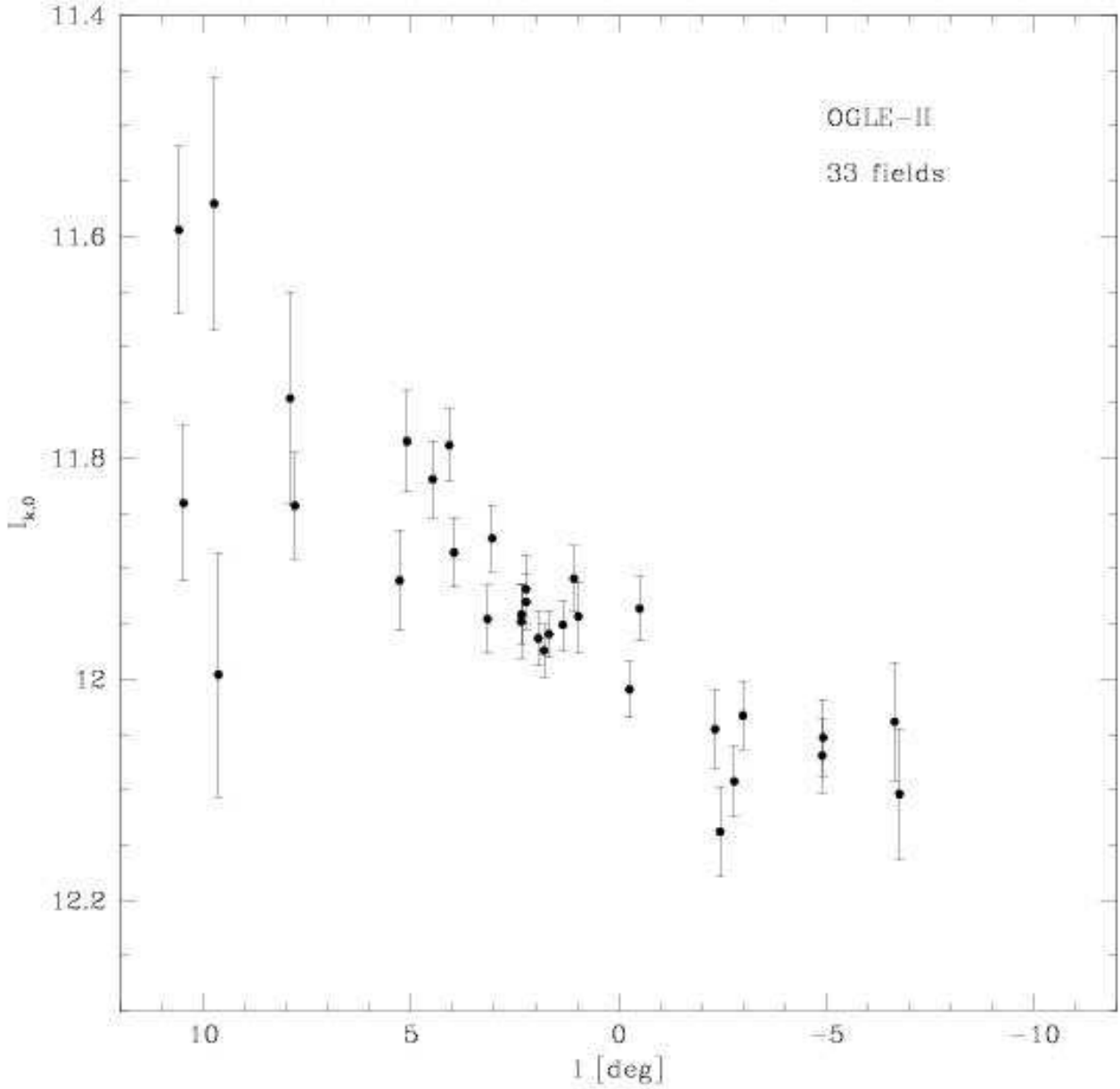}}
\caption{\label{fig4}
  The reddening corrected average I-band magnitude for OSARG type A variables
  in 33 OGLE-II fields is shown as a function of Galactic longitude $l$.
}
\end{figure*}

Fig.~\ref{fig4} presents the variation of the average magnitude of
OSARG type A variables with the Galactic longitude.  A clear trend is
present, practically identical to that shown by the red clump giants
(\citealt{stanek94}, \citealt{stanek97}, \citealt{sumi03}),
demonstrating that OSARG variables are located in the Galactic Bar.

We repeated the analysis for OSARG type B variables and it was
qualitatively similar to that shown in Fig.~\ref{fig4}, but disturbed
by the OGLE-II saturation limit.

\section{Multiple periods}
\begin{figure*}
\resizebox{\hsize}{!}{\includegraphics[angle=0]{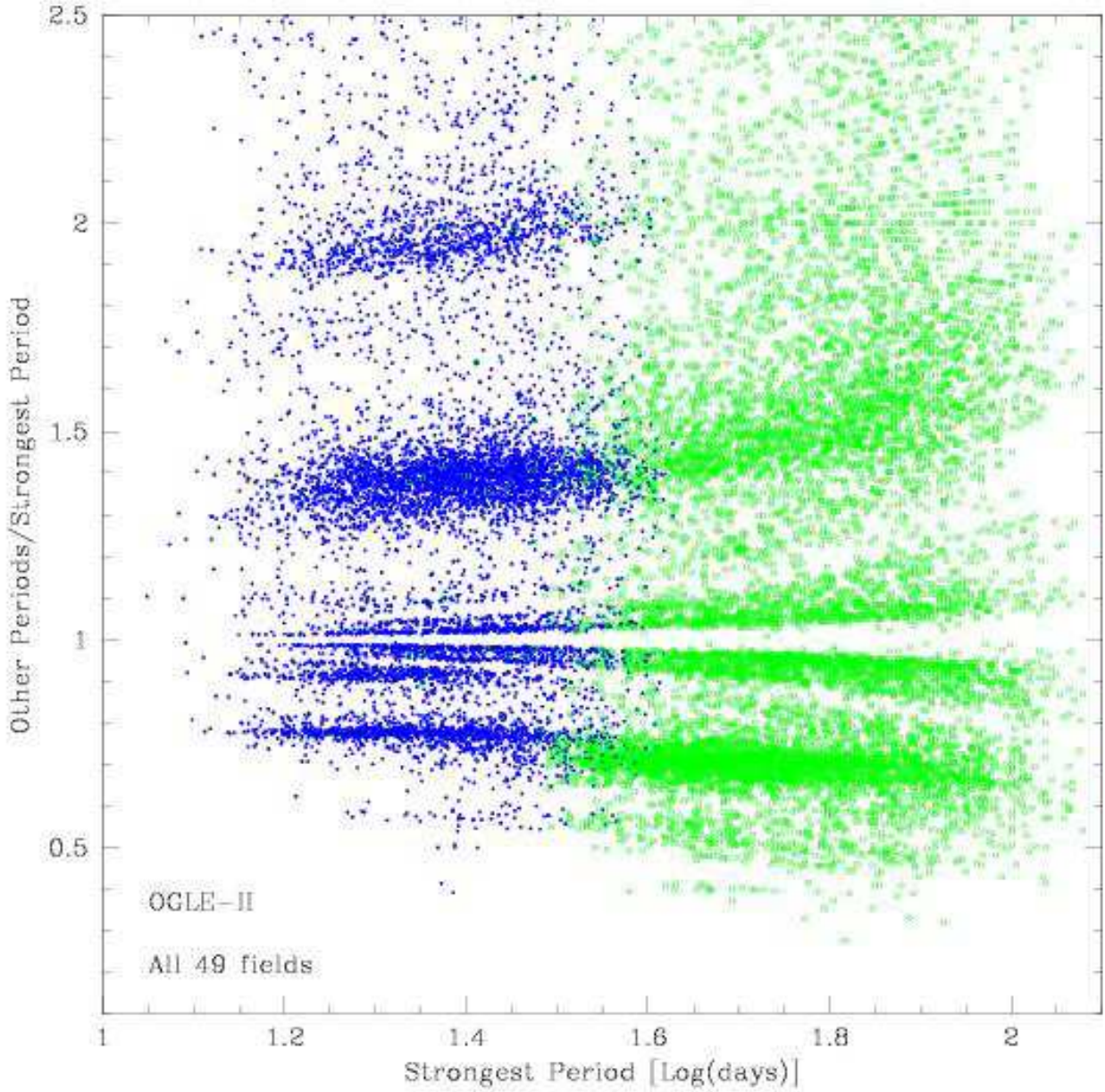}}
\caption{\label{fig5}
  The ratios of all significant periods to the dominant period are shown as
  a function of the dominant period for all OSARG variables, type A (blue
  points) and type B (green open circles).
}
\end{figure*}

Multiple periods were detected for most OSARGs.  The ratios of those
periods to the dominant period (i.e., the one with the largest
amplitude) are shown in Fig.~\ref{fig5} as a function of the dominant
period.  Several bands corresponding to the period ratios close to 2,
1.4, 0.9, 0.75, and 0.5 are clearly seen.  The interpretation of this
rich spectrum of oscillations is beyond the scope of our paper.  Our
intent is to present this richness with a hope that it may attract
specialists to investigate the nature of the oscillations.

Virtually all OSARGs were identified with objects in the 2MASS
catalogue, and the values of J, H, and K magnitude are included in the
tables available on Internet/Simbad. A sample of
these tables is shown in our Table~\ref{tab2}.  These magnitudes were
corrected for interstellar extinction using maps developed by
\citet{sumi03} and using extinction ratios between the infrared bands and
V-band provided by \citet{schlegel98}.  Note that the extinction in
K-band is small, and any errors in the estimate are of relatively little
consequence.  The average K-band magnitude, corrected for extinction,
varies with Galactic longitude in a way similar to that seen in
Fig.~\ref{fig4}.

\begin{figure*}
\resizebox{\hsize}{!}{\includegraphics[angle=0]{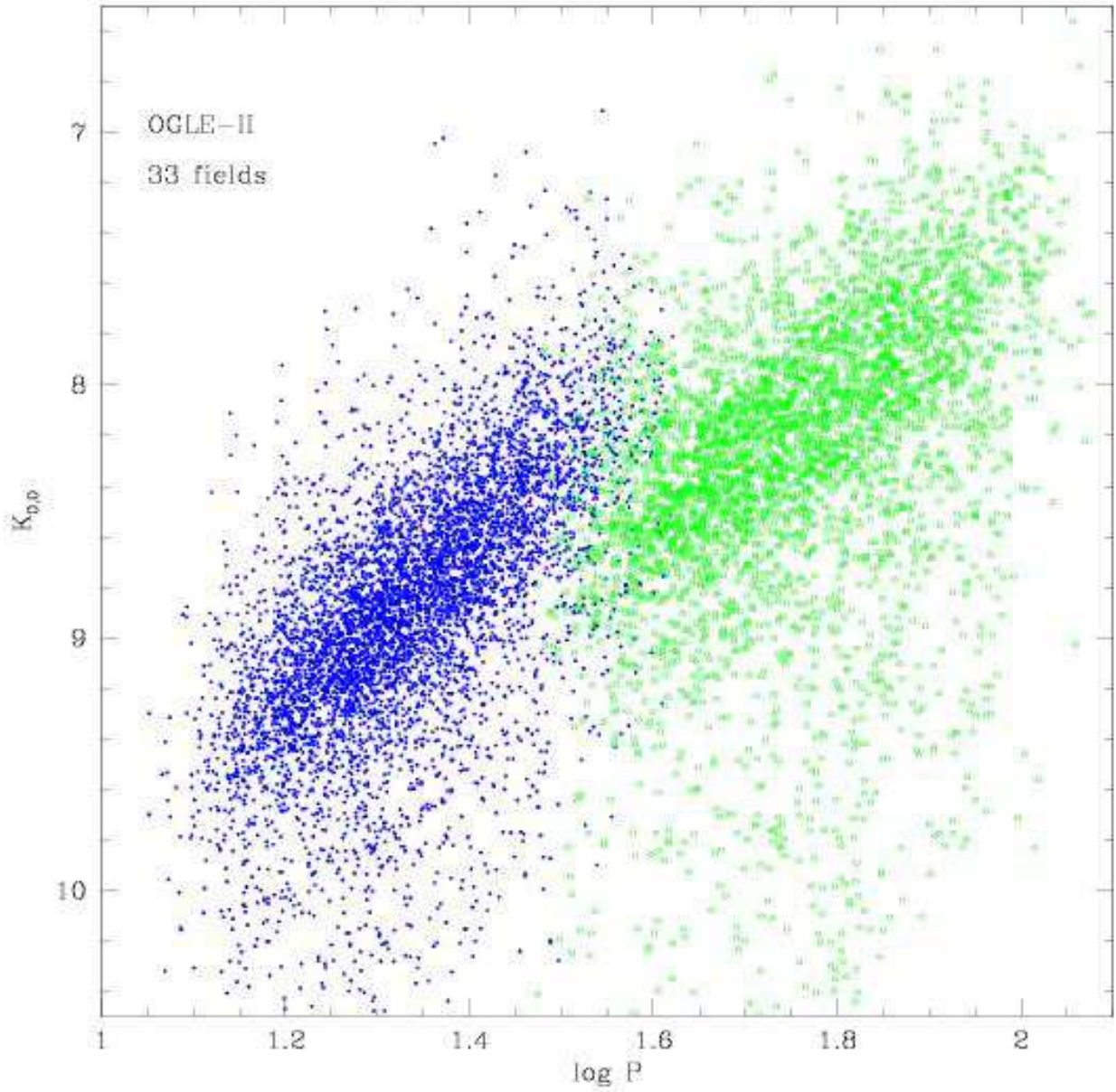}}
\caption{\label{fig6}
  The period -- luminosity diagram for OSARGs in K-band.
  The K-band magnitude of all OSARG variables was obtained from 2MASS and
  corrected for the interstellar extinction and the difference in distance
  induced by the Galactic Bar (cf. Fig. 4 of this paper), yielding
  $ K_{0,0} $, which is shown as a function of the dominant period.  Note
  the clear separation of the two types of OSARG variables, type A (blue
  points) and type B (green open circles).
}
\end{figure*}

Next, extinction corrected $ K_0 $ magnitudes were also corrected for the
Galactic Bar effect (as shown in Fig.~\ref{fig4} of this paper) to the
distance corresponding to Galactic longitude $ l = 0 $; we label the
corresponding magnitudes $ K_{0,0} $.  A relation between the dominant
period of oscillations and $ K_{0,0} $ is shown in Fig.~\ref{fig6}.  The two
groups of OSARGs that are visible in Fig.~\ref{fig1} are also clearly
visible in Fig.~\ref{fig6}.

\begin{figure*}
\resizebox{\hsize}{!}{\includegraphics[angle=0]{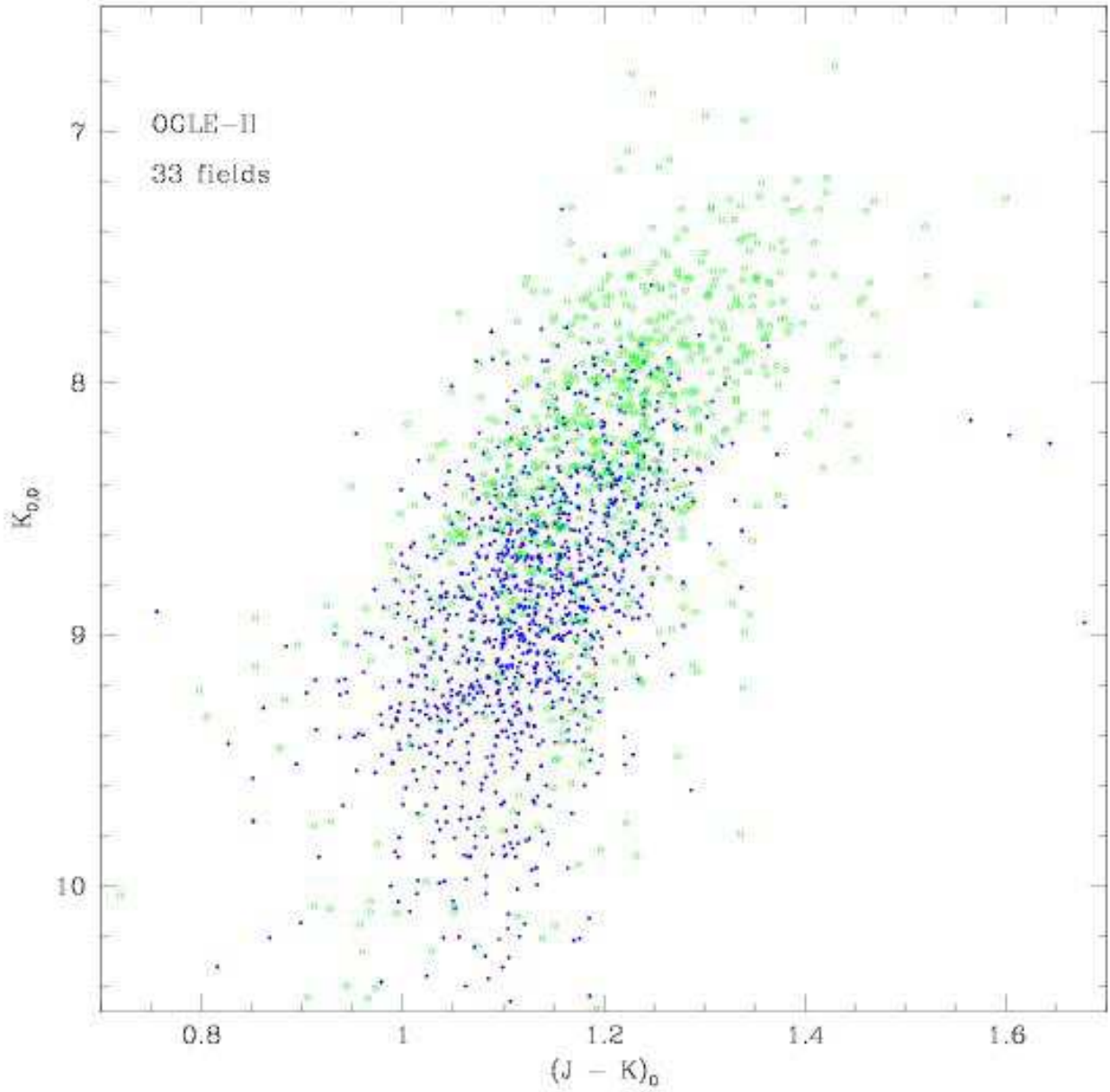}}
\caption{\label{fig7}
  The infrared colour -- magnitude diagram for OSARGs based on 2MASS J and K
  magnitudes corrected for the interstellar extinction and, in the case of 
  $ K_{0,0} $, also for the Galactic Bar-induced difference in distance,
  as determined from Fig. 4 of this paper.  Only 20\% of all variables
  are plotted to avoid crowding.  Notice that type A OSARGs (blue points)
  and type B OSARGs (green open circles) overlap in some area of the CMD.
}
\end{figure*}

The two groups of OSARGs are also plotted in the infrared color --
magnitude diagram in Fig.~\ref{fig7}. Only 20\% of all variables are
plotted to avoid crowding.  There is almost full overlap between the
two groups, distinctly different from the clear separation in
Fig.~\ref{fig1} and Fig.~\ref{fig6}.

\begin{figure*}
\resizebox{\hsize}{!}{\includegraphics[angle=0]{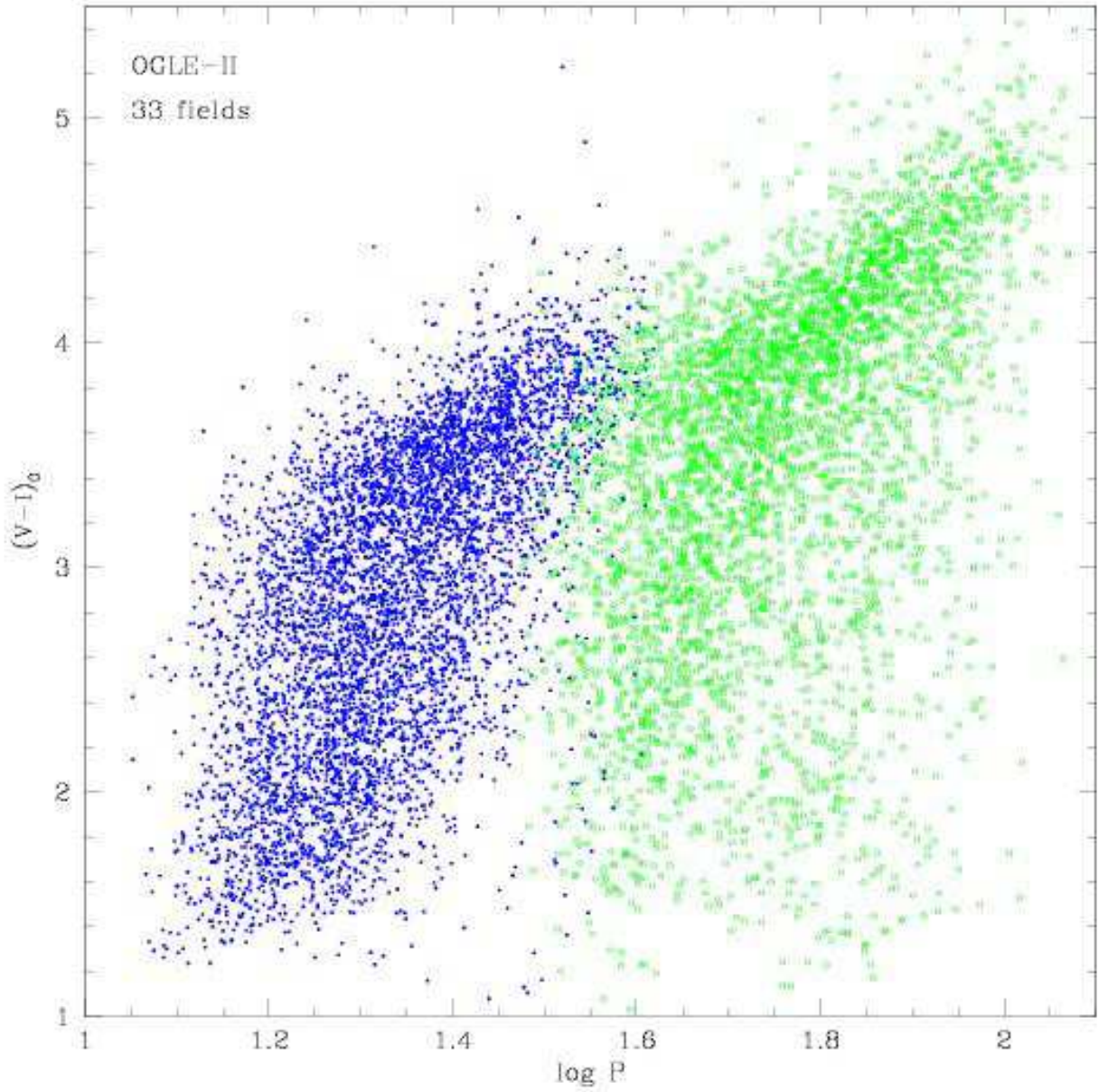}}
\caption{\label{fig8}  
  The reddening corrected $ (V-I)_0 $ colour as a function of the
  dominant period. Two distinct period -- colour relations, for type A
  (blue points) and for type B (green open circles), are clearly apparent.
}
\end{figure*}

Fig.~\ref{fig8} shows the dependence of the dominant period on the
reddening corrected $ (V-I)_0 $ color, and clear period -- color
relations are apparent.  Note: K-band magnitude is close to the
bolometric magnitude, and hence the period -- luminosity relations are
clearly visible in Fig.~\ref{fig6}.  However, the bright giants form
a nearly horizontal line in the $ I_0 - (V-I)_0 $ diagram shown in
Fig.~\ref{fig2}.  Hence, the $ (V-I)_0 $ color is for them a better
absolute luminosity indicator than the $ I_0 $ magnitude, and this is
clearly demonstrated in Fig.~\ref{fig8}.

\begin{figure*}
\resizebox{\hsize}{!}{\includegraphics[angle=0]{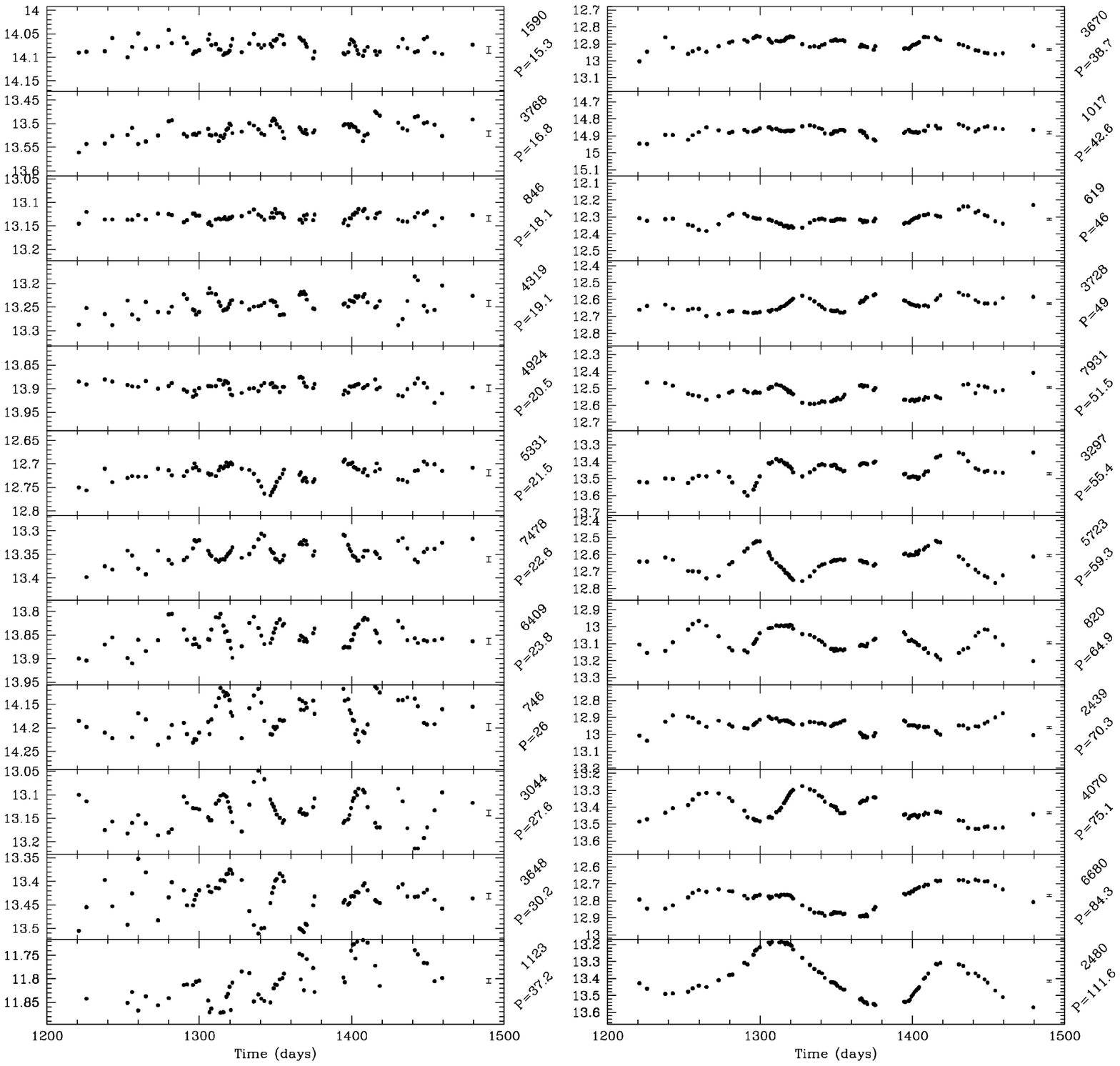}}
\caption{\label{fig9}
  A sample of light curves with periods (in days) increasing
  from top to bottom, for type A (left) and type B (right) variables.
}
\end{figure*}

Finally, it is time for us to present a random sample of type A and type B
OSARGs in Fig.~\ref{fig9}. These are 12 variables of each
type in the field ${\rm BUL\_SC3}$, sorted with increasing dominant period.
These appear as semi-regular variables because of their rich period
structure.  A much longer time baseline of photometric measurements
will become available soon when the current OGLE-III data are
processed and published (cf. Udalski et al. 2002a).  With a longer time
baseline, it will be much easier to decide to what extent the multiple
periods we found are stable, and to what extent OSARGs are
multiply-periodic rather than semi-regular variables.

\section{The Catalogue}

We provide a catalogue of 8970 OSARG type A variables and 6399 OSARG
type B variables in electronic form at:

\centerline{http://www.astro.princeton.edu/$\sim$jwray/OSARG/}
\noindent
in the files A.txt and B.txt, respectively.  Separate files for each
field are available at the same URL, and are called A1.txt, A2.txt,
\ldots, A49.txt and B1.txt, B2.txt, \ldots, B49.txt.  The catalogue is
a subset of the general OGLE-II catalogue of variable stars in the
Galactic Bulge \citep{wozniak01}.

\noindent
A sample of the tables, listing data for the first 25 type A stars from the
field ${\rm BUL\_SC1}$, is shown in Table~\ref{tab2}, which is split into
two parts because of its considerable width.  The consecutive columns are
as follows: the variable star number according
to \citet{wozniak01}; the star number according to OGLE maps
(\citealt{udalski02b}); 2000 RA and DEC; X and Y pixel coordinates;
average V-band magnitude; V-band extinction; average (V-I) color;
(V-I) reddening; average I-band magnitude; I-band extinction; J, H, K-band
magnitudes; the dominant period and its amplitude; the next three
significant periods and amplitudes.  In cases where fewer than three
significant periods were found, the extra columns contain the value
-99.99999.  Similarly, for a few stars there was no match found in the 2MASS
catalogue, so the J, H, K columns for those stars contain the value -9.999.

The coordinates and I and V-band magnitudes are given following OGLE
maps (\citealt{udalski02b}), the reddening and extinction is given
following \citet{sumi03}, and the J, H, K magnitudes are given following
2MASS (All-Sky Point Source Catalogue, Release 2003 March 25).  We provide
all these data in electronically accessible files to facilitate analysis
of OSARG variables to all interested users.

\section{Discussion}

There can be no doubt that OSARGs have been found in the MACHO database
for the Galactic Bulge \citep{cook97}, and that they are related to
red giant variables found in the LMC and SMC by \citet{cook97}, by
\citet{kiss03} and by \citet{ita03}.  However, there is also a striking
difference between the Galactic Bulge and the LMC and SMC: the
multiple narrow bands seen in the $ \log P - K $ diagrams for the LMC and
SMC are not present in the Galactic Bulge, where we see only two
groups of stars.  It appears that our type A and B variables correspond
to type ${\rm A^-}$ and ${\rm B^-}$ variables, respectively, of Ita et al.
(2003, fig.~6).  Note: the distance modulus to the Galactic
Bulge is 4 mag smaller than that to the LMC, so the red giant tip, which is
at ${\rm K \approx 12}$ in the LMC, is expected to be at ${\rm K \approx 8 }$
in the Galactic Bulge.  Unfortunately, this is close to the I-band
saturation limit of OGLE, and this complicates the comparison and
biases our diagrams.

We have an impression that our two sequences of variables are broader
in the apparent magnitude than the corresponding sequences seen in the LMC
and SMC.  This may be caused by at least two effects: the Galactic Bar has
a relatively large radial depth, which remains an effect even after
correction is made for the inclination (cf. Fig.~\ref{fig4}), and
there is likely also a broad range of metallicities in our Galaxy as
compared to LMC and SMC.

Another striking difference between OSARGs and similar variables in
the LMC is the presence of a large number of LMC stars with color
${\rm (J-K) > 1.4 }$, as shown in fig.~3 of \citet{kiss03}; these
are carbon stars.  We found very few such variables, as is apparent
in our Fig.~\ref{fig7}, where there are less than a handful of stars off the
right hand limit of the figure.  This difference may be due to the
higher metallicity of the Galactic Bulge as compared with the LMC.

The period -- amplitude relations as presented in our Fig.~\ref{fig1}
have not been reported in print to the best of our knowledge.
However, similar relations were found for the LMC and SMC variables
by \citet{soszynski03}.

We restrict this paper to the announcement of the OSARG variables in
the Galactic Bulge/Bar, to the catalogue of 15,369 objects, and to the
demonstration that they are located in a bar inclined to the line of
sight.  We make no attempt at modelling the Bar, as considerably more
complete data should be available as soon as ASAS photometry (cf.
\citealt{pojmanski02}) becomes available, and the problems with the
OGLE saturation limit are alleviated.

\section*{Acknowledgments}
This work was supported with the NSF grant AST-0204908, and NASA grant
NAG5-12212.  This publication makes use of data products from the Two
Micron All Sky Survey, which is a joint project of the University of
Massachusetts and the Infrared Processing and Analysis
Center/California Institute of Technology, funded by the National
Aeronautics and Space Administration and the National Science
Foundation

\newpage
\begin{table*}
 \caption{\label{tab1}
   Values for 33 fields of the reddening corrected
   average magnitude $I_{k,0}$ and its rms, as well as
   the Galactic longitude $l$ and latitude $b$ of the
   field centers, and the number of A and B stars, $N_A$, $N_B$.
 }
  \begin{center}
  \begin{tabular}{rrrrrrr} \hline
Field $k$ &  $l$  & $b$   & $N_A$& $N_B$ & $I_{k,0}$ & rms \\ \hline
1         &  1.08 & -3.62 &  225 & 136   &     11.91 & 0.36\\       
2         &  2.23 & -3.46 &  204 & 133   &     11.92 & 0.41\\       
8         & 10.48 & -3.78 &   39 &  32   &     11.84 & 0.52\\       
9         & 10.59 & -3.98 &   34 &  27   &     11.59 & 0.66\\       
10        &  9.64 & -3.44 &   16 &  17   &     12.00 & 0.62\\       
11        &  9.74 & -3.64 &   15 &  20   &     11.57 & 0.45\\       
12        &  7.80 & -3.37 &   84 &  55   &     11.84 & 0.55\\       
13        &  7.91 & -3.58 &   21 &  22   &     11.75 & 0.47\\       
16        &  5.10 & -3.29 &   92 &  92   &     11.79 & 0.47\\       
17        &  5.28 & -3.45 &   93 &  91   &     11.91 & 0.48\\       
18        &  3.97 & -3.14 &  197 & 108   &     11.89 & 0.46\\       
19        &  4.08 & -3.35 &  181 & 121   &     11.79 & 0.45\\       
20        &  1.68 & -2.47 &  422 & 257   &     11.96 & 0.43\\       
21        &  1.80 & -2.66 &  337 & 179   &     11.97 & 0.42\\       
22        & -0.26 & -2.95 &  294 & 197   &     12.01 & 0.41\\       
23        & -0.50 & -3.36 &  224 & 175   &     11.94 & 0.42\\       
24        & -2.44 & -3.36 &  120 & 150   &     12.14 & 0.43\\       
25        & -2.32 & -3.56 &  154 & 109   &     12.05 & 0.40\\       
26        & -4.90 & -3.37 &  174 & 113   &     12.07 & 0.35\\       
27        & -4.92 & -3.65 &  164 & 102   &     12.05 & 0.43\\       
28        & -6.76 & -4.42 &   57 &  40   &     12.10 & 0.40\\       
29        & -6.64 & -4.62 &   68 &  46   &     12.04 & 0.41\\       
30        &  1.94 & -2.84 &  321 & 196   &     11.96 & 0.40\\       
31        &  2.23 & -2.94 &  313 & 157   &     11.93 & 0.43\\       
32        &  2.34 & -3.14 &  261 & 137   &     11.94 & 0.40\\       
33        &  2.35 & -3.66 &  174 & 116   &     11.95 & 0.40\\       
34        &  1.35 & -2.40 &  387 & 275   &     11.95 & 0.43\\       
35        &  3.05 & -3.00 &  216 & 146   &     11.87 & 0.41\\       
36        &  3.16 & -3.20 &  214 & 138   &     11.95 & 0.42\\       
38        &  0.97 & -3.42 &  195 & 160   &     11.94 & 0.41\\       
40        & -2.99 & -3.14 &  214 & 154   &     12.03 & 0.43\\       
41        & -2.78 & -3.27 &  196 & 174   &     12.09 & 0.35\\       
42        &  4.48 & -3.38 &  158 & 130   &     11.82 & 0.47\\ \hline
 \end{tabular}
 \end{center}
\end{table*}

\begin{table*}
\caption{\label{tab2}
     Table of all relevant parameters for a sample of type A stars in field 1.}
\begin{center}
\scriptsize
  \begin{tabular}{rrrrrrrrrrrr} \hline
NoV & NoS    & RA          & DEC         & X-p     & Y-p    & V      & A(V)  & (V-I) & E(V-I) & I     & A(I)  \\ \hline
7   & 43     & 18:02:09.55 & -30:25:14.8 &  311.41 &  97.78 & 17.640 & 1.558 & 4.436 & 0.793 & 13.195 & 0.765 \\ 
17  & 189902 & 18:02:20.02 & -30:25:44.1 &  638.85 &  27.45 & 15.906 & 1.612 & 3.973 & 0.820 & 11.926 & 0.791 \\ 
67  & 47     & 18:02:04.36 & -30:24:37.1 &  148.87 & 188.64 & 16.774 & 1.720 & 3.570 & 0.876 & 13.199 & 0.844 \\ 
91  & 189911 & 18:02:23.19 & -30:24:45.9 &  737.63 & 168.23 & 17.812 & 1.746 & 4.944 & 0.889 & 12.859 & 0.857 \\ 
136 & 554673 & 18:02:52.20 & -30:24:36.9 & 1644.51 & 191.09 & 16.042 & 1.595 & 2.906 & 0.812 & 13.130 & 0.783 \\ 
149 & 19     & 18:02:03.70 & -30:23:44.5 &  127.99 & 315.63 & 14.878 & 1.700 & 2.571 & 0.865 & 12.304 & 0.834 \\ 
154 & 15     & 18:02:08.06 & -30:24:08.8 &  264.47 & 257.18 & 15.898 & 1.731 & 2.873 & 0.881 & 13.020 & 0.850 \\ 
176 & 189919 & 18:02:29.47 & -30:23:59.4 &  933.64 & 280.90 & 16.641 & 1.569 & 3.710 & 0.799 & 12.924 & 0.770 \\ 
206 & 554654 & 18:02:50.62 & -30:23:48.5 & 1594.97 & 307.91 & 17.204 & 1.552 & 4.524 & 0.790 & 12.672 & 0.762 \\ 
212 & 554680 & 18:03:00.55 & -30:24:05.9 & 1905.53 & 266.15 & 17.325 & 1.708 & 4.225 & 0.870 & 13.091 & 0.839 \\ 
226 & 66     & 18:02:04.59 & -30:22:54.4 &  155.33 & 436.61 & 16.938 & 1.637 & 3.511 & 0.833 & 13.421 & 0.804 \\ 
230 & 29     & 18:02:07.99 & -30:22:39.8 &  261.73 & 472.18 & 16.046 & 1.628 & 3.630 & 0.829 & 12.410 & 0.799 \\ 
251 & 189976 & 18:02:31.04 & -30:22:45.9 &  982.43 & 458.63 & 15.787 & 1.521 & 2.614 & 0.775 & 13.169 & 0.747 \\ 
261 & 376353 & 18:02:35.57 & -30:22:33.2 & 1123.94 & 489.42 & 16.818 & 1.489 & 4.165 & 0.758 & 12.645 & 0.731 \\ 
312 & 201645 & 18:02:19.74 & -30:22:06.3 &  628.98 & 553.78 & 15.425 & 1.547 & 2.822 & 0.787 & 12.598 & 0.759 \\ 
313 & 201647 & 18:02:20.07 & -30:21:38.4 &  639.17 & 621.36 & 16.116 & 1.718 & 3.210 & 0.875 & 12.901 & 0.844 \\ 
328 & 387446 & 18:02:41.92 & -30:21:38.7 & 1322.35 & 621.36 & 17.307 & 1.556 & 4.577 & 0.792 & 12.722 & 0.764 \\ 
343 & 565509 & 18:02:52.41 & -30:21:54.6 & 1650.68 & 583.25 & 15.576 & 1.728 & 3.126 & 0.879 & 12.443 & 0.848 \\ 
344 & 565507 & 18:02:52.69 & -30:21:57.7 & 1659.37 & 575.75 & 15.713 & 1.728 & 2.848 & 0.879 & 12.860 & 0.848 \\ 
368 & 11929  & 18:02:06.27 & -30:20:46.1 &  207.04 & 746.77 & 16.164 & 1.779 & 3.430 & 0.906 & 12.728 & 0.874 \\ 
395 & 201653 & 18:02:18.83 & -30:20:59.4 &  600.22 & 715.45 & 14.756 & 1.681 & 2.491 & 0.856 & 12.260 & 0.825 \\ 
403 & 201655 & 18:02:31.71 & -30:20:53.9 & 1002.91 & 729.26 & 16.301 & 1.517 & 3.735 & 0.772 & 12.560 & 0.745 \\ 
438 & 565514 & 18:03:00.60 & -30:20:59.3 & 1906.77 & 717.13 & 15.725 & 1.595 & 3.669 & 0.812 & 12.048 & 0.783 \\ 
454 & 201665 & 18:02:16.25 & -30:19:51.0 &  518.93 & 880.53 & 15.038 & 1.706 & 2.543 & 0.868 & 12.490 & 0.837 \\ 
463 & 201663 & 18:02:20.67 & -30:20:00.5 &  657.37 & 857.82 & 15.540 & 1.711 & 2.870 & 0.871 & 12.664 & 0.840 \\ 
\end{tabular}
\end{center}

\begin{center}
\scriptsize

\begin{tabular}{rrrrrrrrrrrr} \hline
NoV & J      & H     & K     & Pd.1    & Amp.1   & Pd.2     & Amp.2     & Pd.3     & Amp.3     & Pd.4     & Amp.4      \\ \hline
7   & 10.629 & 9.613 & 9.216 & 19.9844 & 0.01798 & -99.9999 & -99.99999 & -99.9999 & -99.99999 & -99.9999 & -99.99999  \\ 
17  &  9.711 & 8.686 & 8.296 & 32.5068 & 0.03572 & -99.9999 & -99.99999 & -99.9999 & -99.99999 & -99.9999 & -99.99999  \\ 
67  & 11.097 &10.041 & 9.649 & 20.6378 & 0.00881 & -99.9999 & -99.99999 & -99.9999 & -99.99999 & -99.9999 & -99.99999  \\ 
91  &  9.730 & 8.688 & 8.196 & 36.7674 & 0.04686 & 504.9388 & 0.03138   &  50.4939 &   0.02320 & 291.3108 &   0.02110  \\ 
136 & 11.247 &10.222 & 9.937 & 20.8652 & 0.00998 &  27.1472 & 0.00978   &  20.1976 &   0.00635 &  59.6384 &   0.00597  \\ 
149 & 10.523 & 9.561 & 9.261 & 17.2138 & 0.00727 & 236.6901 & 0.00572   & -99.9999 & -99.99999 & -99.9999 & -99.99999  \\ 
154 & 11.165 &10.141 & 9.776 & 13.3582 & 0.00566 &1262.3469 & 0.00537   & -99.9999 & -99.99999 & -99.9999 & -99.99999  \\ 
176 & 10.737 & 9.760 & 9.350 & 18.9827 & 0.01322 & -99.9999 & -99.99999 & -99.9999 & -99.99999 & -99.9999 & -99.99999  \\ 
206 & 10.001 & 8.953 & 8.538 & 29.8192 & 0.04676 &  51.5244 & 0.02070   & 504.9388 &   0.01790 &  75.7408 &   0.01323  \\ 
212 & 10.587 & 9.544 & 9.129 & 20.4153 & 0.01291 & -99.9999 & -99.99999 & -99.9999 & -99.99999 & -99.9999 & -99.99999  \\ 
226 & 11.308 &10.305 & 9.906 & 18.5185 & 0.01553 & 315.5867 & 0.00866   & -99.9999 & -99.99999 & -99.9999 & -99.99999  \\ 
230 & 10.294 & 9.291 & 8.924 & 22.8135 & 0.01736 &  33.0746 & 0.01281   &  46.7536 &   0.01207 & 541.0058 &   0.01264  \\ 
251 & 11.433 &10.475 &10.160 & 14.2907 & 0.00574 &  27.2449 & 0.00533   & -99.9999 & -99.99999 & -99.9999 & -99.99999  \\ 
261 & 10.095 & 9.103 & 8.696 & 23.2259 & 0.02700 &  22.7332 & 0.01737   & -99.9999 & -99.99999 & -99.9999 & -99.99999  \\ 
312 & 10.768 & 9.750 & 9.426 & 20.3604 & 0.01763 &  29.1311 & 0.00832   &  18.5639 &   0.00644 & -99.9999 & -99.99999  \\ 
313 & 10.920 & 9.894 & 9.532 & 20.5260 & 0.00979 &  28.6897 & 0.00775   & -99.9999 & -99.99999 & -99.9999 & -99.99999  \\ 
328 &  9.947 & 8.956 & 8.453 & 28.7988 & 0.02869 & 541.0058 & 0.01892   & -99.9999 & -99.99999 & -99.9999 & -99.99999  \\ 
343 & 10.529 & 9.556 & 9.160 & 15.8454 & 0.00857 &  12.2558 & 0.00615   & -99.9999 & -99.99999 & -99.9999 & -99.99999  \\ 
344 & 11.004 &10.048 & 9.727 & 14.9982 & 0.00580 & -99.9999 & -99.99999 & -99.9999 & -99.99999 & -99.9999 & -99.99999  \\ 
368 & 10.675 & 9.631 & 9.260 & 23.8179 & 0.01882 & 473.3801 & 0.01516   & 176.1414 &   0.01160 &  34.7435 &   0.01090  \\ 
395 & 10.585 & 9.598 & 9.250 & 16.5373 & 0.00668 & 118.3450 & 0.00578   & -99.9999 & -99.99999 & -99.9999 & -99.99999  \\ 
403 & 10.374 & 9.295 & 8.879 & 28.5814 & 0.01643 &  40.2877 & 0.01585   & 541.0059 &   0.01578 & -99.9999 & -99.99999  \\ 
438 &  9.933 & 8.933 & 8.575 & 26.4828 & 0.01878 & 445.5342 & 0.01378   &  38.8414 &   0.01258 &  27.7439 &   0.01009  \\ 
454 & 10.750 & 9.733 & 9.440 & 17.8634 & 0.00683 &  34.5848 & 0.00617   & -99.9999 & -99.99999 & -99.9999 & -99.99999  \\ 
463 & 10.782 & 9.756 & 9.314 & 18.5185 & 0.00890 & 582.6217 & 0.00719   &  24.0447 &   0.00534 & -99.9999 & -99.99999  \\ 
\end{tabular}
\end{center}
\end{table*}

\label{lastpage}

\clearpage


\begin{thebibliography}{}

\bibitem[\protect\citeauthoryear{Blitz \& Spergel}{1991}]{blitz91}
    Blitz L., Spergel D. N., 1991, ApJ, 379, 631
\bibitem[\protect\citeauthoryear{Christensen-Dalsgaard et al.}{2001}]%
    {christensen01}
    Christensen-Dalsgaard J., Kjeldsen H., Mattei J. A., 2001, ApJ, 562, L141
\bibitem[\protect\citeauthoryear{Cook et al.}{1997}]{cook97}
    Cook K. H. et al. (MACHO), 1997, in Ferlet R., 
    Maillard J.-P., Raban B., eds,
    12th IAP Astrophysics Meeting,
    Variable stars and the astrophysical returns of microlensing surveys. 
    Editions Fronti\`eres, Gif-sur-Yvette
    p.~17
\bibitem[\protect\citeauthoryear{Dehnen}{2002}]{dehnen02}
    Dehnen W., 2002,
    in Athanassoula E., Bosma A., Mujica R., eds,
    ASP Conf. Ser. 275, 
    Disks of Galaxies:
    Kinematics, Dynamics and Perturbations.
    Astron. Soc. Pac., San Francisco,
    p.~105
\bibitem[\protect\citeauthoryear{de Vaucouleurs}{1964}]{devaucouleurs64}
    de Vaucouleurs G., 1964,
    in  Kerr F. J., Rodgers A. W., eds,
    Proc. IAU Symp. 20,
    The Galaxy and the Magellanic Clouds.
    (Canberra: Australian Acad. Sci.), 
    p.~195
\bibitem[\protect\citeauthoryear{Dziembowski et al.}{2001}]{dziembowski01}
    Dziembowski W. A. et al., 2001, MNRAS, 328, 601
\bibitem[\protect\citeauthoryear{Eyer \& Grenon}{1997}]{eyer97}
    Eyer L., Grenon M., 1997, ESA SP-402, 467
\bibitem[\protect\citeauthoryear{Eyer \& Blake}{2002}]{eyer02}
    Eyer L., Blake C., 2002, 
    in  Aerts C., Bedding T. R., Christensen-Dalsgaard J., eds,
    ASP Conf. Ser. 259,
    Radial and Nonradial Pulsations as Probes of Stellar Physics.
    Astron. Soc. Pac., San Francisco,
    p.~160
\bibitem[\protect\citeauthoryear{Garz\'on}{1999}]{garzon99}
    Garz\'on F., 1999,
    in Beckman J. E., Mahoney T. J., eds,
    ASP Conf. Ser. 187,
    The Evolution of Galaxies on Cosmological Timescales.
    Sheridan Books, San Francisco, 
    p.~31
\bibitem[\protect\citeauthoryear{Gerhard}{2002}]{gerhard02}
    Gerhard O., 2002,
    in Da Costa G. S., Jerjen H., eds,
    ASP Conf. Ser. Vol. 273, 
    The Dynamics, Structure and History of Galaxies:
    A Workshop in Honour of Prof. K. Freeman.
    Astron. Soc. Pac., San Francisco,
    p.~73
\bibitem[\protect\citeauthoryear{Glass \& Schultheis}{2003}]{glass03}
    Glass I. S., Schultheis M., 2003, preprint (astro-ph/0307366)
\bibitem[\protect\citeauthoryear{Ita et al.}{2003}]{ita03}
    Ita Y. et al., 2003, preprint (astro-ph/0310083)
\bibitem[\protect\citeauthoryear{Jorissen et al.}{1997}]{jorissen97}
    Jorissen A. et al., 1997, A\&A, 324, 578
\bibitem[\protect\citeauthoryear{Kiss \& Bedding}{2003}]{kiss03}
    Kiss L. L., Bedding T., 2003, MNRAS, 343, L79
\bibitem[\protect\citeauthoryear{Lomb}{1976}]{lomb76}
    Lomb N. R., 1976, Ap\&SS, 39, 447
\bibitem[\protect\citeauthoryear{Matsumoto et al.}{1982}]{matsumoto82}
    Matsumoto T. et al., 1982,
    in Riegler G., Blandford R., eds,
    The Galactic Center.
    American Institute of Physics, New York,
    p.~48
\bibitem[\protect\citeauthoryear{Merrifield}{2003}]{merrifield03}
    Merrifield M. R., 2003, preprint (astro-ph/0308302)
\bibitem[\protect\citeauthoryear{Nakada et al.}{1991}]{nakada91}
    Nakada Y. et al., 1991, Nature, 353, 140
\bibitem[\protect\citeauthoryear{Pojma\'nski}{2002}]{pojmanski02}
    Pojma\'nski G., 2002, AcA, 52, 397
\bibitem[\protect\citeauthoryear{Press et al.}{1992}]{press92}
    Press W. H. et al., 1992, Numerical recipes in Fortran.
    Cambridge University Press
\bibitem[\protect\citeauthoryear{Schlegel et al.}{1998}]{schlegel98}
    Schlegel D., Finkbeiner D. P., Davis M., 1998, ApJ, 500, 525
\bibitem[\protect\citeauthoryear{Soszy\'nski}{2003}]{soszynski03}
    Soszy\'nski I. (OGLE), 2003, PhD Thesis, Warsaw University
\bibitem[\protect\citeauthoryear{Stanek et al.}{1994}]{stanek94}
    Stanek K. Z. et al. (OGLE), 1994, ApJ, 429, L73
\bibitem[\protect\citeauthoryear{Stanek et al.}{1997}]{stanek97}
    Stanek K. Z. et al. (OGLE), 1997, ApJ, 477, 163
\bibitem[\protect\citeauthoryear{Sumi}{2003}]{sumi03}
    Sumi T., 2003, preprint (astro-ph/0309206)
\bibitem[\protect\citeauthoryear{Udalski et al.}{1992, OGLE--I}]{udalski92}
    Udalski A. et al. (OGLE), 1992, AcA, 42, 253
\bibitem[\protect\citeauthoryear{Udalski et al.}{1997, OGLE--II}]{udalski97}
    Udalski A., Kubiak M., Szyma\'nski M. (OGLE), 1997, AcA, 47, 319
\bibitem[\protect\citeauthoryear{Udalski et al.}{2002a, OGLE--III}]{udalski02a}
    Udalski A. et al. (OGLE), 2002a, AcA, 52, 1
\bibitem[\protect\citeauthoryear{Udalski et al.}{2002b}]{udalski02b}
    Udalski A. et al. (OGLE), 2002b, AcA, 52, 217
\bibitem[\protect\citeauthoryear{Wood et al.}{1999}]{wood99}
    Wood P. R. et al. (MACHO), 1999,
    in Le Bertre T., Lebre A., Waelkens C., eds,
    Proc. IAU Symp. 191, 
    Asymptotic giant branch stars.
    Astron. Soc. Pac., San Francisco,
    p.~151
\bibitem[\protect\citeauthoryear{Wood}{2000}]{wood00}
    Wood P. R., 2000, Publ. Astr. Soc. Australia, 18, 18
\bibitem[\protect\citeauthoryear{Wo\'zniak et al.}{2001}]{wozniak01}
    Wo\'zniak P. R. et al. (OGLE), 2001, AcA, 51, 175

\end{thebibliography}
\end{document}